\documentclass[aps,twocolumn,amsfonts,amsmath,amssymb,superscriptaddress,10pt,floatfix,nofootinbib]{revtex4-1}
\usepackage{times}
\usepackage[english]{babel}
\usepackage[utf8]{inputenc}

\usepackage{amsmath}
\usepackage{amsfonts}
\usepackage{amssymb}
\usepackage{amsthm}
\usepackage{microtype}
\usepackage{graphics}
\usepackage[dvips]{graphicx}
\usepackage{color}
\usepackage{rotating}
\usepackage[normalem]{ulem}
\usepackage {ulem}
\usepackage{import}

\begin{document}

\title{Molecular mechanism for cavitation in water under tension}
\author{Georg Menzl}
\affiliation{Faculty of Physics and Center for Computational Materials Science, University of Vienna, Boltzmanngasse 5, 1090 Vienna, Austria}

\author{Miguel A. Gonzalez}
\affiliation{Department of Chemistry, Imperial College London, London SW7 2AZ, United Kingdom}

\author{Philipp Geiger}
\affiliation{Faculty of Physics and Center for Computational Materials Science, University of Vienna, Boltzmanngasse 5, 1090 Vienna, Austria}

\author{Frédéric Caupin}
\affiliation{Institut Lumière Matière, UMR5306 Université Claude Bernard Lyon 1 - CNRS, Université de Lyon, Institut Universitaire de France, 69622 Villeurbanne cedex, France} 

\author{Jose L. F. Abascal}
\affiliation{Departamento de Qu\'{\i}mica F\'{\i}sica, Facultad de Ciencias Qu\'{\i}micas, Universidad Complutense de Madrid, 28040 Madrid, Spain}

\author{Chantal Valeriani}
\affiliation{Departamento de Qu\'{\i}mica F\'{\i}sica, Facultad de Ciencias Qu\'{\i}micas, Universidad Complutense de Madrid, 28040 Madrid, Spain}
\affiliation{Departamento de Fisica Aplicada I, Facultad de Ciencias Fisica, Universidad Complutense de Madrid, 28040 Madrid, Spain}

\author{Christoph Dellago}
\email[Corresponding author: ]{christoph.dellago@univie.ac.at}
\affiliation{Faculty of Physics and Center for Computational Materials Science, University of Vienna, Boltzmanngasse 5, 1090 Vienna, Austria}

\begin{abstract}
Despite its relevance in biology and engineering, the molecular mechanism driving cavitation in water remains unknown.
Using computer simulations, 
we investigate the structure and dynamics of vapor bubbles emerging from metastable water at negative pressures.
We find that in the early stages of cavitation, bubbles are irregularly shaped and become more spherical as they grow.
Nevertheless, the free energy of bubble formation can be perfectly
reproduced in the framework of classical nucleation theory (CNT) if the curvature dependence of the surface tension is taken into account.
Comparison of the observed bubble dynamics to the predictions of the macroscopic Rayleigh--Plesset (RP) equation, augmented with thermal fluctuations, demonstrates that the growth of nanoscale bubbles is governed by viscous forces.
Combining 
the dynamical prefactor determined from the RP equation with 
CNT based on Kramers' formalism yields an analytical expression for the cavitation rate that reproduces the simulation results very well over a wide range of pressures. Furthermore, our theoretical predictions are in excellent agreement with cavitation rates obtained from inclusion experiments.
This suggests that homogeneous nucleation is observed in inclusions, whereas only heterogeneous nucleation on impurities or defects occurs in other experiments.
\end{abstract}

\maketitle

Due to its pronounced cohesion, water remains stable under tension for long times.
Experimentally, strongly negative pressures exceeding $-120\, {\rm MPa}$  \cite{GreenAngellScience1990,ZhengAngellScience1991,AlvarengaBodnarJCP1993,AzouziCaupinNaturePhys2013,MercuryShmulovichNATO2014,PallaresCaupinPNAS2014} can be sustained before the system decays into the vapor phase via cavitation, i.e., bubble nucleation. 
Recently, cavitation in water under tension has drawn research interest due to its importance in biological processes, like water transport in natural \cite{StroockHolbrookRevFluidMech2014,PonomarenkoMarmottantRoyalSoc2014,LarterDelzonPlantPhys2015,RowlandOliveiraNature2015} and synthetic \cite{WheelerStroockNature2008,VincentOhlPRL2012} trees, spore propagation of ferns \cite{NoblinDumaisScience2012}, and poration of cell membranes \cite{OhlLohseBiophysJ2006,AdhikariBerkowitzJPhysChemB2015}.
Furthermore, cavitation in water appears to be the driving force behind the sonocrystallization of ice~\cite{OhsakaTrinhAppPhysLett1998,YuWangUltraSonochem2012} and preventing its occurrence remains a challenge in turbine and propeller design~\cite{Kumar2010}.
Studying the onset of cavitation has also proven to be a valuable tool to locate the line of density maxima in metastable water \cite{AzouziCaupinNaturePhys2013}, which contributes to the ongoing effort of explaining the origin of water's anomalies \cite{PallaresCaupinPNAS2014,DebenedettiNaturePhys2013}.
Interest in the topic is magnified by the startling discrepancy arising when cavitation in water is investigated using different experimental methods. 
While agreement between different methods is excellent in the high-temperature regime, where the liquid is unable to sustain large tension, a significantly higher degree of metastability is reached when studying cavitation in inclusions along an isochoric path \cite{GreenAngellScience1990,ZhengAngellScience1991,AlvarengaBodnarJCP1993,AzouziCaupinNaturePhys2013,MercuryShmulovichNATO2014} compared to other techniques \cite{HerbertCaupinPRE2006,DavittCaupinEPL2010} at low temperatures \cite{CaupinHerbertCRPhys2006}. 

Due to the short time-scale on which the transition takes place and the small volume of the critical bubble at experimentally feasible conditions, direct observation of cavitation at the microscopic level remains elusive. However, cavitation rates are directly accessible in experiment and 
some microscopic insight into
the cavitation transition can be obtained from these data by means of the nucleation theorem \cite{KashchievJCP1982}, which relates the variation in the height of the free energy barrier separating the metastable liquid from the vapor phase upon change of external parameters to properties of the critical bubble \cite{AzouziCaupinNaturePhys2013,DavittCaupinEPL2010}. The microscopic information that can be inferred is limited and, since not all quantities entering the nucleation theorem are known, {\it ad hoc} assumptions have to be introduced.
For state-points where cavitation is a rare event, classical nucleation theory (CNT) can be invoked to provide a qualitative understanding of the transition \cite{CaupinPRE2005}.
However, while CNT provides a physically meaningful and appealingly simple picture of nucleation processes, the estimates for the nucleation rates obtained from CNT are known to differ substantially (up to many orders of magnitude) from those measured in experiments 
\cite{CaupinHerbertCRPhys2006,ZengOxtobyJCP1991,OxtobyJPhysCondMatter1992}.

Computer simulations are a natural choice to investigate cavitation in water with molecular resolution on the time-scales governing the emergence of microscopic bubbles in the liquid. While cavitation in simple liquids has been studied extensively using computer simulations \cite{shen99, VishnyakovNeimarkPRE2000, NeimarkVishnyakovJCP2005, WangFrenkelJPhysChemB2009, BaisdakovTeterinJCP2011, meadley12, torabi13b}, simulation studies of cavitation in water were focused on methodological aspects \cite{AbascalValerianiJCP2013, ourfirstpaper, GonzalezBresmeJCP2015} or performed at state points in vicinity of the vapor--liquid spinodal \cite{ZahnPRL2004, ChoKimPRL2014}. 
In this work, we apply a combination of several complementary computer simulation methods to identify the molecular mechanism of cavitation. 
A statistical committor analysis carried out on reactive trajectories reveals that the volume of the largest bubble in the system
constitutes a good reaction coordinate for bubble nucleation.
We compute the dynamics of nanoscale bubbles along this reaction coordinate and demonstrate that the pressure dependence of the bubble diffusivity can be reproduced by Rayleigh--Plesset (RP) theory generalized to include thermal fluctuations, thereby elucidating the crucial influence of viscous damping on bubble growth.
Based on Kramers' formalism and the RP equation we obtain an analytical expression for the nucleation rate that yields excellent agreement with numerical results obtained for a wide range of pressures with a method akin to the Bennett--Chandler approach for the computation of reaction rate constants.
The obtained rates are validated for selected points by comparison to estimates from transition interface sampling and 
support estimates obtained from inclusion experiments. To augment the microscopic picture of cavitation we characterize the morphology of bubbles in water under tension and analyze the bubble surface in terms of its hydrogen bonding structure.

\section{Classical nucleation theory}
Our investigations are guided by CNT, which posits that the decay of the metastable liquid under tension proceeds via the formation of a small vapor bubble, whose growth is initially opposed by a free energy barrier.
According to Kramers' theory \cite{KramersPhysica1940,SchultenSzaboJCP1981}, the escape rate $k$ from a well over a high barrier for a system moving diffusively in a potential $U(q)$ along a coordinate $q$ is given by
$k = \left[ \left( \int_\cup \exp[- \beta U(q)] {\rm d}q \right) \left( \int_\cap \exp [\beta U(q) ] / D(q) {\rm d}q \right) \right]^{-1}.$
Here, the symbols $\cup$ and $\cap$ indicate that the integration is carried out over the well and the barrier, respectively, and $D(q)$ is the diffusion coefficient. 
In order to describe bubble nucleation, we use the volume $v$ of the largest bubble in the system as the order parameter (committor calculations \cite{DellagoGeissler2002} indicate that $v$ is indeed a good reaction coordinate, see Appendix) and we replace the potential energy by the potential of mean force $-k_{\rm B}T \ln [v_0 P(v)]$, where $k_{\rm B}$ ist the Boltzmann constant, $P(v)$ is the probability density that the largest bubble is of size $v$, and $v_0$ is an arbitrary constant volume.
Assuming that the diffusion coefficient does not change appreciably on the top of the barrier and approximating the barrier to second order, one obtains the nucleation rate (number of nucleation events per unit time and unit volume)
\begin{equation}\label{eq:rates2}
 J = \frac{\omega D(v^\ast)}{\sqrt{2 \pi k_B T}} \frac{P(v^\ast)}{V},
\end{equation}
where $v^\ast$ is the critical bubble volume, $V$ is the total volume of the system, and $\omega$ is related to the barrier curvature $\kappa$ by $\omega^2 = -\kappa$.
This functional form provides a physical picture of the waiting time associated with (rare) transitions by factorizing the rate $J$ into a kinetic part $\propto \omega D(v^\ast)$ and the probability density $P(v^\ast)$ of encountering a bubble with volume $v^\ast$, i.e., a configuration that relaxes to the vapor or the liquid phase with equal probability.
In the following, we will compute the probability $P(v^\ast)$ to find a bubble of critical size
and derive an analytical expression for the diffusion constant $D(v^\ast)$ needed in the CNT rate expression.

\section{Free energy of cavitation at negative pressures}
Using umbrella sampling simulations, we have computed the equilibrium bubble density $\rho(v)$ at a temperature $T = 296.4 \, {\rm K}$ and various pressures (see Methods). 
For large bubbles, $\rho(v)$ is equal to the probability density $P(v)/V$ for the volume of the largest bubble as needed in Eq.~(\ref{eq:rates2})
\cite{MaibaumPRL2008}.
The equilibrium bubble density is related to the Gibbs free energy $g(v)$ of a bubble of volume $v$ by $g(v) = -k_{\rm B} T \ln [ \rho(v)/\rho_0 ]$, where $\rho_0$ is a constant included to make the argument of the logarithm dimensionless. The value of  $\rho_0$ is fixed by requiring that the Gibbs free energy of a bubble of size $v = 0$ vanishes.
Note that the constant $\rho_0$, required to relate the cavitation free energy $g(v)$ to the equilibrium bubble density $\rho(v)$, is not specified in the framework of CNT. Various choices for $\rho_0$ have been made in the literature without rigorous justification, as discussed in the Methods Section. Here, we use information from molecular simulations to determine the value of $\rho_0$ unambiguously (see Appendix).  

\begin{figure}[t]
\centering
 \includegraphics[width=0.35\textwidth]{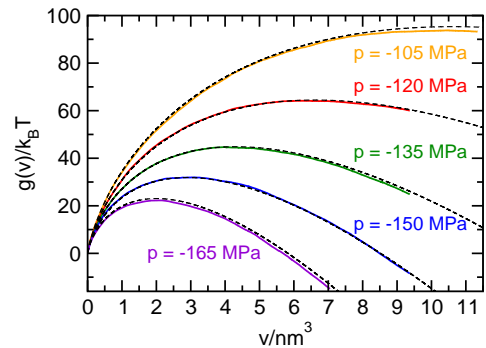}
 \caption{
 Free energy $g(v)$ of bubble nucleation as a function of bubble volume $v$ for various negative pressures at $T = 296.4 \, {\rm K}$ obtained from umbrella sampling calculations. Dashed lines indicate CNT-predictions from Eq.~(\ref{eq:tolman}), which describe the free energy very accurately over the investigated pressure range. In the framework of CNT, the curves can be understood as a result of the competition between the free-energetic cost of forming the liquid--vapor interface (which dominates in the case of small bubbles) and the mechanical work gained from expanding the system under tension (favoring large bubbles).
 The location of the resulting maximum in the free energy corresponds to the volume of the critical bubble $v^\ast$: bubbles of this volume are least likely to be encountered in an equilibrium configuration and overcoming this free energy barrier is the rate-limiting step in cavitation away from the spinodal. 
 \label{fig:fvonv}}
\end{figure}

We obtain a quantitative description of the cavitation free energy within CNT by examining the free energetic cost of the bubble interface, i.e., the free energy without the mechanical work $p v$ gained from expanding the system under tension, per surface area (see Appendix). 
Remarkably, the free energetic cost of the vapor--liquid interface is independent of pressure within the accuracy of our computations and as such, for the wide range of pressures investigated, the free energy of cavitation differs only by the mechanical work $p v$.
We find that CNT describes the free energy of bubble nucleation accurately, provided that the curvature dependence of the surface tension $\gamma$ is taken into account. In particular, the free energy of cavitation is reproduced by
\begin{equation} \label{eq:tolman}
 g(v) =  4 \pi r^2(v) \frac{\gamma_0}{1 + 2 \delta / r(v)}  + pv \,,
\end{equation}
where $r(v) = (3 v/4 \pi)^{1/3}$ is the radius of a sphere with volume $v$.
Here, the parameters $\gamma_0 = 20.24 \, k_{\rm B}T/{\rm nm^2}$ and $\delta = 0.195 \, {\rm nm}$ are obtained from a fit to the free energetic cost of the liquid--vapor interface. 
Bubble free energies $g(v)$ for various pressures as well as the estimates from Eq.~(\ref{eq:tolman}), which agree almost perfectly with the simulation data (dashed black lines),
are shown in Fig.~\ref{fig:fvonv}.
Over the range of bubble volumes studied here, the value of $\delta$ obtained from the fit is positive, which indicates that the concave curvature of the interface decreases the surface tension $\gamma$, thereby favoring bubbles over droplets (a discussion of the curvature dependence of the surface tension is provided in the Appendix).

\section{Bubble morphology}
\begin{figure}[t]
\centering
\includegraphics[width=0.36\textwidth]{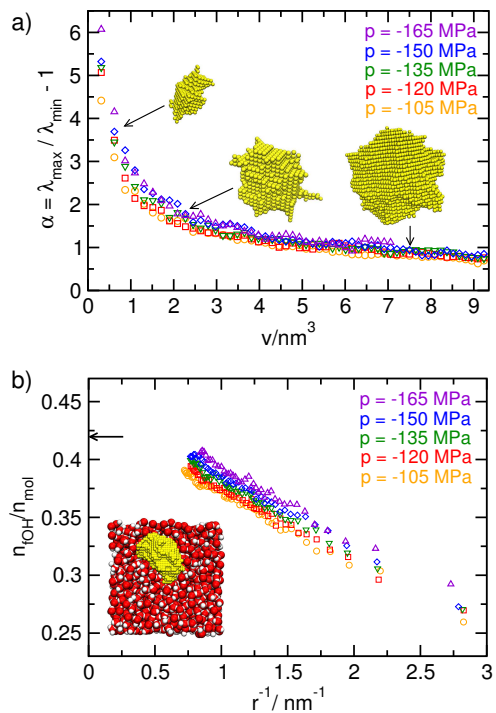}
\caption{
Shape and hydrogen bonding structure of bubbles.
{\it a)} Asphericity $\alpha$ as a function of bubble volume from configurations obtained via umbrella sampling. 
By construction, $\alpha$ is zero for a perfect sphere and higher values indicate shapes with higher aspect ratios.
The inset shows bubbles (not to scale) observed at $p = -150 \, {\rm MPa}$ whose asphericities and volumes are indicated by arrows.
{\it b)} Fraction of free OH groups $n_{\rm fOH}/n_{\rm mol}$ at the bubble surface as a function of the inverse radius $r^{-1}$ of a sphere with volume $v$. 
The arrow indicates the fraction $n_{\rm fOH}/n_{\rm mol}$ for a flat interface at $300\, {\rm K}$ at ambient pressure from Ref.~\cite{VilaverdeCampenJPhysChemB2012}. Note that we give the fraction of broken hydrogen bonds per molecule, so the highest possible value is 2.
The depicted configuration contains a bubble of critical size at a pressure of $p = -150 \, {\rm MPa}$, where the yellow spheres indicate the unoccupied grid-points forming the largest bubble.
\label{fig:combofig}}
\end{figure}

At the conditions studied here, bubbles are essentially voids in the metastable liquid which, for bubble volumes $v \lesssim 10 \, {\rm nm^3}$, rarely contain vapor molecules \cite{AbascalValerianiJCP2013, ourfirstpaper}. Visual inspection indicates that small bubbles mostly have an irregular shape which becomes more compact as the bubbles grow larger (some representative bubbles of different size are depicted in Fig.~\ref{fig:combofig}a). 
Larger bubbles are predominantly compact and may be viewed as resembling spheres with strongly undulating surfaces \cite{AbascalValerianiJCP2013,ourfirstpaper}. This observation is confirmed by computing the average asphericity of bubbles defined as $\alpha = \lambda_{\rm max}/\lambda_{\rm min} - 1$, where $\lambda_{\rm max}$ and $\lambda_{\rm min}$ are the largest and smallest eigenvalue of the gyration tensor of the bubble, respectively. As shown in Fig.~\ref{fig:combofig}a, the asphericity is only weakly dependent on pressure and decreases with increasing bubble volume.

The free energetic cost of forming bubbles in water is intimately connected to breaking and re-arranging hydrogen bonds (HBs) at the interface. The hydrogen bonding structure at the liquid--vapor interface
depends on the size of the bubble \cite{LumWeeksJPhysChemB1999,ChandlerNature2005}. For small bubbles, HBs in the liquid are re-arranged and the fraction of broken HBs at the interface is similar to that of the bulk liquid whereas in the case of large bubbles, the bubble surface becomes similar to the flat vapor--liquid interface.
As shown in Fig. \ref{fig:combofig}b, the number of broken HBs per molecule at the interface increases with bubble size and
the fraction of free OH groups at the interface decays roughly linearly with its mean curvature $r^{-1}$ over the studied range of bubble volumes.

\section{Bubble dynamics}
Since CNT with a curvature dependent surface tension describes the free energy of cavitation very accurately, thus providing the volume $v^\ast$ of the critical bubble and the curvature $-\omega^2$ of the barrier, all that is needed to predict rates via Eq.~(\ref{eq:rates2}) is the diffusivity $D(v^\ast)$ of the bubble volume in the barrier region.
In the following, we use the Rayleigh--Plesset (RP) equation \cite{RayleighPlesset,Kagan,LeightonUltrasonics2008}, which describes the dynamics of a vapor bubble in a fluid at the macroscopic level, to derive an analytical expression that relates the microscopic diffusion constant $D(v^\ast)$ to the macroscopic properties of the liquid.

The RP equation is the equation of motion for the volume $v$ of a spherical bubble evolving with internal pressure $p_{\rm b}$ in a liquid with mass density $m$, viscosity $\eta$, and surface tension $\gamma$:
\begin{equation}\label{eq:rp}
  m \ddot{v} - \frac{m \dot{v}^2}{6v} = 4 \pi \left( \frac{3 v}{4 \pi }\right)^{\frac{1}{3}} \left[ p_{\rm b} - p - 2 \gamma \left( \frac{4 \pi}{3 v} \right)^{\frac{1}{3}} - \frac{4 \eta }{3} \frac{\dot{v}}{v} \right] .
\end{equation}
Here, for simplicity we neglect the curvature dependence of the surface tension, but stress that the following derivation can be easily generalized (see Appendix) and all results shown in the figures were obtained including this correction. Neglecting the inertial terms on the left hand side of the RP equation, one finds
\begin{equation}
  \dot{v} = - \frac{3 v}{4 \eta} \left[ p + 2 \gamma \left( \frac{4 \pi}{3 v} \right)^{\frac{1}{3}} \right] = - \frac{1}{\Gamma(v)} \frac{{\rm d}g(v)}{{\rm d} v} \, ,
\end{equation}
where we assumed that the pressure inside the bubble is negligible. In the above equation we have rewritten the right hand side in order to indicate that the time evolution of the volume $v$ can be viewed as an overdamped motion on the CNT free energy $g(v) = (36 \pi v^2)^{1/3} \gamma + p v$ under the effect of the friction $\Gamma(v) = 4 \eta / 3 v$.

Since for microscopic bubbles thermal fluctuations play an important role, the RP equation is augmented with a random force $F(t) = \sqrt{2 k_{\rm B} T / \Gamma(v)} \xi(t)$, where $\xi(t)$ is Gaussian white noise and the magnitude of the force is determined by the fluctuation--dissipation theorem. The diffusion coefficient for the bubble volume then follows from the Einstein relation,
$D(v) = 3 k_{\rm B} T v / 4 \eta$ (note that this result holds also if the surface tension depends on the mean curvature of the bubble). Inserting the critical $v^\ast = 32 \pi \gamma^3/3 \vert p \vert^3$ we finally obtain the diffusion coefficient at the top of the barrier needed for the rate calculation
\begin{equation}\label{eq:DRP}
 D(v^\ast) = \frac{8 \pi k_{\rm B} T \gamma^3}{\eta \vert p \vert^3} \, .
\end{equation}
Including the curvature dependence of the surface tension for $v^\ast$ and $\gamma$ yields a similar, but slightly more complicated formula (see Appendix).
 
\begin{figure}[t]
\centering
 \includegraphics[width=0.35\textwidth]{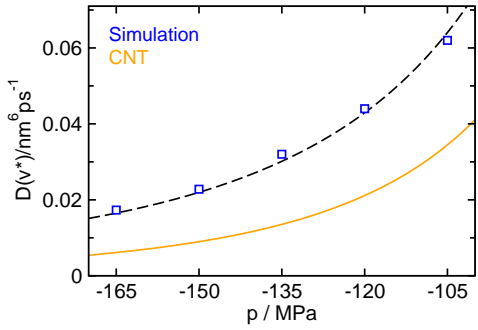}
 \caption{
 The diffusion constant $D(v^\ast)$ on top of the free energy barrier obtained from the Rayleigh--Plesset equation predicts the correct scaling with pressure $p$.
The RP estimate (orange line) was obtained by using the volume $v^\ast$ of the critical bubble and the curvature $-\omega^2$ of the barrier from CNT, including a curvature dependent surface tension.
The scaling behavior of the diffusion constant obtained from simulation (blue squares) is illustrated by a fit $\propto p^{-3}$ (dashed black line).
\label{fig:D}}
\end{figure}
A comparison between the diffusion constant $D(v^\ast)$ obtained from the RP-equation combined with CNT and the estimate obtained directly from simulation (see Methods) is shown in Fig.~\ref{fig:D}. The viscosity at negative pressures needed in the formula for the diffusion constant was determined in molecular dynamics simulations using the Green--Kubo relation (see Appendix). 
The analytical formula obtained from the RP-CNT approach underestimates the diffusivity in comparison to simulation results only by about a factor of two, which is remarkable considering that this estimate is obtained from a macroscopic approach based on hydrodynamics. Moreover, by virtue of the pressure dependence of $v^\ast$ in CNT, it predicts the scaling $\propto \vert p \vert^{-3}$ of the diffusion constant with pressure accurately, suggesting that the dynamics of bubble growth are essentially controlled by the viscosity of the liquid.

\section{Cavitation rates}
We are now in a position to predict cavitation rates according to Eq.~(\ref{eq:rates2}) over a wide range of pressures, including the strongest tensions observed in experiment.
As a point of comparison, we have computed cavitation rates numerically using a method akin to the divided-saddle method \cite{DaruStirlingJCTC2014} based on the Bennett--Chandler (BC) \cite{BennettChandler1,BennettChandler2} approach and transition interface sampling (TIS) \cite{ErpBolhuisJCP2003}, respectively (see Methods).
\begin{figure}[tb]
\centering
 \includegraphics[width=0.35\textwidth]{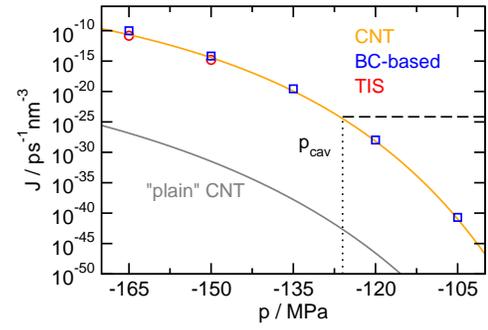}
 \caption{
Predictions obtained from CNT using microscopic information are in excellent agreement with cavitation rates $J$ from direct simulation.
The estimates obtained from simulations by a variant of the Bennett--Chandler method (blue squares) agree well with the transition interface sampling (red circles) reference calculations (see Methods). Predictions of curvature-corrected CNT (orange line) with the correct value of $\rho_0$ utilizing the kinetic prefactor shown in Fig.~\ref{fig:D} yield excellent agreement with simulation results, while ``plain'' CNT (grey line) severely underestimates the cavitation rate. For ``plain'' CNT, we chose $\rho_0 = n_l n_v$, where $n_l$ and $n_v$ are the number density of the liquid and the vapor, respectively \cite{BlanderKatzAIChE1975}. These rate estimates allow for a direct comparison to conflicting experimental predictions on the stability of water under tension by computing the cavitation pressure $p_{\rm cav}$. Following Ref.~\cite{CaupinHerbertCRPhys2006}, we define $p_{\rm cav}$ such that the probability to observe a cavitation event is $P = 1/2$ in a system of volume $V = 1000\, {\rm \mu m}^3$ over an observation time of $\tau = 1 \, {\rm s}$. Assuming that the cavitation events are associated with an exponential waiting time, as is typical for activated processes, a rate of $J = \ln 2/(V \tau)$ (dashed black line) is compatible with this requirement. Its intersection with the CNT prediction gives the cavitation pressure $p_{\rm cav} \approx -126\, {\rm MPa}$.
 \label{fig:rates}}
\end{figure}
The obtained cavitation rates, shown in Fig.~\ref{fig:rates}, vary by more than $30$ orders of magnitude over the studied range of pressures.
The numerical results are accurately reproduced by CNT based on Eq.~(\ref{eq:rates2}) with a curvature-dependent surface tension and the correct value of $\rho_0$ as well as the kinetic prefactor from the RP equation.
In contrast, ``plain'' CNT, i.e., CNT with a constant surface tension and a commonly used expression for $\rho_0$ (see Methods), underestimates the cavitation rates by more than $15$ orders of magnitude.
This shortcoming illustrates the importance of including microscopic information, such as a curvature-dependent surface tension and the correct value of $\rho_0$, for the accurate prediction of rates.

By computing the cavitation pressure $p_{\rm cav}$ from the rates shown in Fig.~\ref{fig:rates} we can directly compare the results obtained here to the conflicting experimental estimates for the limit of metastability of water under tension.
The obtained estimate for the cavitation pressure $p_{\rm cav} \approx -126\, {\rm MPa}$ is in line with 
the results obtained in inclusion experiments \cite{GreenAngellScience1990,ZhengAngellScience1991,AlvarengaBodnarJCP1993,AzouziCaupinNaturePhys2013,MercuryShmulovichNATO2014,PallaresCaupinPNAS2014}. In contrast, the predicted cavitation tension is more negative by about $100 \, {\rm MPa}$ than the data obtained via other experimental techniques would suggest \cite{HerbertCaupinPRE2006,CaupinHerbertCRPhys2006}. Since the simulation setup excludes impurities in the fluid by design, this suggests that cavitation in these cases is indeed heterogeneous as was suspected in previous works \cite{AzouziCaupinNaturePhys2013,DavittCaupinEPL2010}, which explains the significantly lower stability of water under tension in these experiments (a detailed discussion is provided in the Appendix).

\section{Conclusions}

At ambient temperature and strong tension, bubbles in metastable water are essentially voids in the liquid whose shape can deviate significantly from the assumption of a spherical nucleus made in CNT, depending on their size. 
Nonetheless, provided the dependence of the surface tension on the average curvature is included, the free energetics of bubble formation can be quantitatively described in the framework of CNT. 
We find that the curvature contribution favors the cavity over the droplet, i.e., $\delta > 0$, in agreement with experimental results \cite{AzouziCaupinNaturePhys2013}.
In light of conflicting results on the sign of $\delta$ in water, further study is required to elucidate the influence of the chosen water model and biasing towards certain cavity shapes on the obtained value of $\delta$.

By including the effect of thermal fluctuations in the Rayleigh--Plesset equation, we obtain an estimate for the bubble diffusivity that 
accurately reproduces the pressure dependence found in simulation and scales inversely with the viscosity of the liquid.
Combining the kinetic pre-factor determined for this diffusivity with the equilibrium bubble density yields a CNT expression for the cavitation rate that reproduces the nucleation rates very well for negative pressures.
However, the microscopic mechanism for cavitation is expected to change for higher pressures and temperatures, where the 
saturated vapor density is significantly higher than at the temperature studied here. At those conditions, similarly to droplet nucleation \cite{BeckerDoeringAnnalPhys1935}, the transport of molecules across the interface via evaporation and condensation will have a stronger influence on the kinetics of bubble growth, thereby diminishing the influence of viscous damping on the dynamics of the bubble.

The estimate for the cavitation pressure obtained from our rate calculations agrees well with the data from inclusion experiments, thus calling the conflicting results harvested by other techniques into question. 
Since the latter methods greatly underestimate the stability of water under tension, heterogeneous cavitation due to impurities is a likely explanation for this discrepancy.

\section{Methods}
\subsection{Simulation details}
We simulate $N = 2000$ water molecules in the isothermal--isobaric ensemble at a temperature of $T = 296.4 \, {\rm K}$ using the rigid, non-polarisable TIP4P/2005 model~\cite{AbascalVegaJCP2005}, where the long-range interactions are treated with Ewald summation.
The rate computations are carried out using molecular dynamics by integrating the equations of motion with a time step of $2 \, {\rm fs}$ using a time-reversible quaternion based integrator that maintains the rigid geometry of water molecules~\cite{Kamberaj2005}. Constant pressure is ensured by a barostat based on the Andersen approach~\cite{AndersenJCP1980} coupled to a Nos\'e--Hoover thermostat chain
\cite{Tuckerman}.
Equilibrium free energies are computed by use of umbrella sampling (US) 
in conjunction with the hybrid Monte Carlo
(HMC)~\cite{DuaneKennedyPhysLettB1987} scheme. 
Here, we employ a modified version of the Miller integrator~\cite{Miller2002} with a Liouville operator decomposition according to Omelyan~\cite{Omelyan2002}, which reduces fluctuations in the total energy significantly, thereby allowing the use of a time step of $7 \, {\rm fs}$. Each HMC step consists of three MD integration steps, constant pressure was implemented by isotropic volume fluctuations according to the Metropolis criterion
and sampling was 
enhanced by replica exchange moves \cite{GeyerThompsonJAmAstat1995} between neighboring windows. For the direct computation of cavitation rates we employ transition interface sampling (TIS) \cite{ErpBolhuisJCP2003}, where we implemented time reversal and replica exchange moves in addition to shooting moves (described in detail in Refs.~\cite{DellagoGeissler2002,ErpPRL2007,BolhuisJCP2008}).  The probability
histograms for the individual windows in US and TIS were spliced together using a self consistent histogram method \cite{FerrenbergSwendsenPRL1989}.

\subsection{Order parameter}
We study homogeneous bubble nucleation from over-stretched metastable
water using the volume of the largest bubble as a local order parameter.
Estimates for the volume $v$ of each bubble present in the system are obtained by use of the V-method, which was developed to give thermodynamically consistent estimates for the bubble volume \cite{ourfirstpaper} \footnote{Note that the nomenclature was adapted to facilitate readability: $v$/$\xi$ in this work corresponds to $V^{\rm V}_{\rm bubble}/v$ in Ref.~\cite{ourfirstpaper}.}.
The V-method is a grid-based clustering approach to bubble detection \cite{WangFrenkelJPhysChemB2009}, calibrated such that its estimate $v$ for the volume of a bubble corresponds to the average change in system volume due to the presence of such a bubble:
\begin{equation}\label{eq:vmethod1}
 v(\xi) = \frac{\partial}{\partial n} \langle V \rangle_{n(\xi)}\, .
\end{equation}
Here, $\xi$ is the preliminary bubble volume estimate from the grid-based method, i.e., the total volume of all vapor-like grid cubes belonging to the bubble, and $\langle V \rangle_{n(\xi)}$ is the average volume of the system when $n$ bubbles of size $\xi$ are present. As such, $v(\xi)$ corresponds to the average change in system volume when a single bubble of size $\xi$ is added to or removed from the system. For large bubbles, i.e., for bubble volumes where $n(\xi)$ is either zero or one and there are no larger bubbles present in the system, Eq.~(\ref{eq:vmethod1}) becomes
\begin{equation}\label{eq:vmethod}
 v(\xi) = \langle V \rangle_{\xi} - \langle V \rangle,
\end{equation}
where $\langle V \rangle_{\xi}$ is the average volume of the system when the largest bubble is of size $\xi$ and $\langle V \rangle$ is the average volume of the unconstrained metastable liquid at the thermodynamic state point.

On average, since the vapor density in the interior of bubbles is negligible, volume estimates obtained by Eq.~(\ref{eq:vmethod}) are equal to those obtained by computing the equimolar dividing surface between liquid and the largest cavity for each configuration. As a result, the obtained estimates for the bubble volume fulfill the nucleation theorem \cite{KashchievJCP1982}, i.e., $\partial g(v^\ast) / \partial p = v^\ast$, and $pv$ corresponds to the mechanical work gained with respect to the metastable liquid by expanding the system volume at negative pressures. Details on the calibration of the V-method for the state-points investigated in this work are given in the Appendix.

\subsection{Bubble density}
To compute the equilibrium bubble density $\rho(v)$, we first carry out a straightforward molecular dynamics simulation and compute $\langle n(v,\Delta v) \rangle$,  
the average number of bubbles with a volume in a narrow interval $[v, v + \Delta v]$. 
To compute $n(v,\Delta v)$ for larger bubbles which do not form spontaneously on the timescale of the simulation, we carry out umbrella sampling simulations with a bias on the volume of the largest
bubble. The resulting curves are joined, thus yielding $\rho(v) = \langle n(v,\Delta v) \rangle/(\langle V \rangle \Delta v)$ over a wide range of bubble volumes.

\subsection{Detecting hydrogen bonds at the liquid--vapor interface}
We identify molecules as belonging to the bubble surface when they are within
$3.5\,$\AA$\,$
of the bubble.
This cutoff radius is identical to the radius of the exclusion spheres used to determine occupied grid points during the evaluation of the order parameter (for an in-depth description see Ref.~\cite{ourfirstpaper}) and thus all water molecules forming the boundary layer in our bubble detection procedure are part of the interface.
When analyzing whether two water molecules form a hydrogen bond with each other, we employ the criterion used in Ref.~\cite{VilaverdeCampenJPhysChemB2012} in a study of the flat vapor--liquid interface in order to facilitate easy comparison between the obtained results. For molecule $A$ to be considered as donating a hydrogen bond to molecule B, two criteria have to be fulfilled simultaneously: The distance between the oxygens 
$d_{\rm O_A O_B} < 3.5\,$\AA$\,$
and the maximum angle 
${\rm O_A - H \cdots O_B} > 140^{\circ}$. 

\subsection{Rate calculation}\label{met:rates}
We employ a method based on the Bennett--Chandler approach \cite{BennettChandler1, BennettChandler2}
to obtain rates estimates without any assumptions about the dynamics of the bubble in the liquid. 
In addition to the states $A$ (metastable liquid) and $B$ (far enough to the right of the free energy barrier such that the system is committed to transitioning to the vapor phase), we introduce a state $S$ around the dividing surface,
akin to the approach taken in the divided-saddle method~\cite{DaruStirlingJCTC2014}.
An ensemble of trajectories, each $L$ steps long, is generated by propagating checkpoints selected from the region $S$ forward and backward in time. From these trajectories one then computes the time correlation function $C_{AB}(t)$, which is the conditional probability to find the system in $B$ at time $t$ provided it is in $A$ at time zero, 
\begin{equation}
C_{AB}(t) = ( L + 1 ) \left\langle \frac{h_{A}(0) h_{B}(t)}{N_{S}[x(\tau)]}\right\rangle_{G} \frac{\langle h_{S} \rangle}{\langle h_{A} \rangle} \, . 
\end{equation}
Here, $h_{A/B}$ is $1$ when the system is in state $A/B$ and zero else, $N_{S}[x(\tau)]$ is the number of configurations of a trajectory $x(\tau)$ in the saddle domain and $\langle \cdots \rangle_{G}$ denotes an average over the trajectories generated from points in $S$.  The ratio $\langle h_{S} \rangle/\langle h_{A} \rangle$ is the equilibrium probability of finding the system in $S$ relative to the equilibrium probability of state $A$ and it can be determined from the free energy $g(v)$. The transition rate constant $k_{AB}$ is then obtained by computing the numerical derivative ${\rm d}C_{AB}/{\rm d}t$ in the time range where $C_{AB}(t)$ is linear.

Nucleation rates calculated at $p = - 165 \, {\rm MPa}$ and $-150 \, {\rm MPa}$ using transition interface sampling \cite{ErpBolhuisJCP2003} (TIS, red circles in Fig.~\ref{fig:rates}) agree with the estimates of the BC-based approach up to statistical error.
As an additional point of comparison, we used the BC-based approach to compute rates at $T = 280\, {\rm K}$ and $p = -225\, {\rm MPa}$, where nucleation is spontaneous on the time-scale of an unconstrained molecular dynamics simulation starting in the metastable liquid. 
The estimate $J = 3.1 \times 10^{-5} {\rm ps^{-1} nm^{-3}}$ obtained from straight-forward MD simulations in Ref.~\cite{AbascalValerianiJCP2013} agrees well with the BC-based estimate of 
$J = 7.4 \times 10^{-5} {\rm ps^{-1} nm^{-3}}$.

\subsection{Computation of the diffusion constant}
Since the volume of the largest bubble is a good reaction coordinate for the transition, its diffusivity can be computed via mean first passage times \cite{BerezhkovskiiSzaboJPhysChemB2013,LuVandenEijndenJCP2014}.
Assuming that the diffusion coefficient does not change significantly in the barrier region, i.e., $D(v) = D(v^\ast)$, to second order it can be expressed as $D = b^2 \left( 1 - \beta b^2 \omega^2/6 \right)/(2 \langle \tau(b) \rangle),$
where $b$ is the distance of the absorbing boundary from the top of the free energy barrier, approximated by an inverted parabola with curvature $-\omega^2$, and $\langle \tau(b) \rangle$ is the mean first passage time for a given value of $b$.
As a starting point at the top of the barrier we used equilibrium configurations created by umbrella sampling where the system contained a cluster of critical size and drew the particle velocities as well as the thermostat and barostat velocities at random from the appropriate Maxwell-Boltzmann distributions.

\subsection{Plain CNT}
As a point of comparison, we obtain an estimate for the cavitation rates from CNT with a constant surface tension $\gamma_0 = 17.09 \, k_{\rm B}T/{\rm nm^2}$ for TIP4P/2005 water \cite{VegaMiguelJCP2007}. The CNT estimate for the rate is given by 
\begin{equation}
 J = \frac{\sqrt{k_{\rm B} T \gamma_0^3}}{\eta \vert p \vert} \rho_0 e^{- \beta 16 \pi \gamma_0^3 / p^2}.
\end{equation}
The equation above
was obtained from  Equs.~(\ref{eq:rates2}) and (\ref{eq:DRP}), where $\omega = p^2/\sqrt{32 \pi \gamma_0^3}$
and the probability density $P(v)/V = \rho_0 \exp (-\beta g(v^\ast))$.
Here, $g(v^\ast) = 16 \pi \gamma_0^3 / 3 p^2$ and the normalization constant was chosen as $\rho_0 = n_l n_v \approx 4.4 \times 10^{-3} {\rm nm^{-6}}$, where $n_l$ and $n_v$ are the number density of the metastable liquid and the number density of the vapor at coexistence \cite{BlanderKatzAIChE1975}, respectively.
Note that the prefactor $\rho_0$ 
is not uniquely defined in the framework of CNT and various choices have been employed in the literature \cite{ZengOxtobyJCP1991,BlanderKatzAIChE1975,OxtobyJPhysCondMatter1992}.
These choices lead to estimates ranging from $\rho_0 = 9.4 \times 10^{-14}\, {\rm nm^{-6}}$ to $\rho_0 = 2.4 \times 10^8\, {\rm nm^{-6}}$ at $p = -135 \, {\rm MPa}$ (we obtain $\rho_0 = 0.02 \, {\rm nm^{-6}}$ from the simulation data shown in Fig.~\ref{fig:fs}).
The resulting predictions for the cavitation rates underestimate the values determined from simulation by $6-27$ orders of magnitude.

\begin{acknowledgments}
We thank S. Garde, P. Geissler, V. Molinero, A. Patel, A. Tr\"oster, E. Vanden-Eijnden, and S. Venkatari for insightful comments.
The work of G.M., P.G., and C.D. was supported by the Austrian Science Foundation (FWF) under grant P24681-N20 and within the SFB ViCoM (Grant No.~F41).
P.G. also acknowledges financial support from FWF grant P22087-N16 and F.C. from ERC under the European FP7 Grant Agreement 240113.
C.V. acknowledges  financial support from a Marie Curie Integration
Grant 322326-COSAAC-FP7-PEOPLE-CIG-2012 and a Ramon y Cajal tenure track. The team at Madrid acknowledges funding from the MCINNC Grant FIS2013-43209-P.
Calculations were carried out on the Vienna Scientific Cluster (VSC).
\end{acknowledgments}



\appendix
\section{Calibration of the order parameter}
Below, we give a brief description on how the V-method, which is employed in this work to obtain an estimate for the volume of the largest bubble, is parametrized to yield a thermodynamically consistent estimate for the bubble volume. For an in-depth description of the method employed to detect bubbles in the metastable liquid, we refer the reader to Ref.~\cite{ourfirstpaper}.

We employ a grid-based procedure to detect bubbles in the system by clustering grid-points that are not occupied by liquid-like water molecules. The preliminary size estimate $\xi$ for the bubble is the total volume of all vapor-like cubes belonging to the same cluster. Here, we use a grid of $52^3$ points (each of which thus corresponds to a cube with volume $V/52^3$) for a system of $N = 2000$ water molecules. The radius of the exclusion spheres which determines the ``volume'' of each water molecule around its center of mass, was chosen as $r_{\rm S} = 3.35\,$\AA, close to the location of the first minimum in the O--O radial distribution function.

As mentioned in the Methods-section, the V-method is calibrated such that its estimate for the volume of the largest bubble $v$ corresponds to the average change in system volume $V$ due to the presence of a bubble. Since this change in system volume depends on the chosen thermodynamic state point, one needs to determine $v$ as a function of the preliminary grid-based order parameter $\xi$ to obtain the correct calibration at the state point of interest. 
\begin{figure}[h]
\centering
 \includegraphics[width=0.35\textwidth]{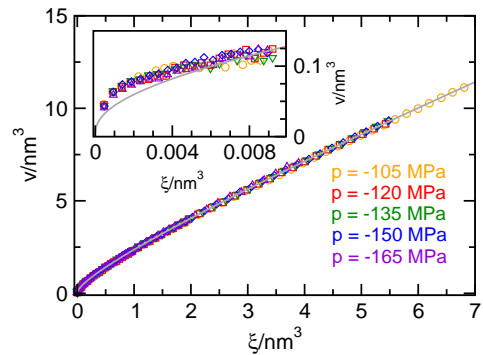}
 \caption{
 Average change $v$ in system volume $V$ due to the presence of a cluster of vapor-like cubes, i.e., a bubble, with volume $\xi$. The grey line indicates the fit given by Eq.~(\ref{eq:fit2}). The inset magnifies the small-bubble regime, where each data point for $v$ was obtained according to Eq.~(\ref{eq:vmethod1}). Data points only shown in the main plot and not included in the inset were obtained according to Eq.~(\ref{eq:vmethod}).
 \label{fig:Vvonv}}
\end{figure}
The data for $T = 296.4 \,{\rm K}$ and various negative pressures is shown in Fig.~\ref{fig:Vvonv}. In order to obtain a convenient mapping of $\xi$ onto $v$ we choose the fitting function, indicated by the grey line in the figure, as
\begin{equation}\label{eq:fit2}
 v(\xi) \approx \xi + k_1 \, \xi^{2/3} + k_2 \, \xi^{1/3}\, .
\end{equation}
Here, $k_1 \approx 1.04 \,{\rm nm}$ and $k_2 \approx 0.33 \,{\rm nm^2}$ produce a mapping that agrees well with the data. The same fitting parameters are used for all pressures shown in the figure, since the data are indistinguishable within the statistical accuracy.

\section{Volume of the largest bubble as a reaction coordinate}

The rate equation of CNT, Eq.~(1), is based on the assumption that the dynamics of bubble growth can be described as the diffusion of a bubble volume on the respective free energy surface.
To quantify to which extent the volume of the largest bubble tracks the progress of the cavitation transition dynamically, i.e., whether the volume of the largest bubble is a reaction coordinate, we perform a statistical committor analysis, which correlates values of the chosen order parameter with the probability $p_{\rm B}$ that the system transitions to the vapor phase.

We create reactive trajectories by propagating equilibrium configurations harvested by means of umbrella sampling close to the size of the critical bubble, which for a pressure of $-150 \, {\rm MPa}$ is $v^\ast = 2.95 \, {\rm nm^3}$, backward and forward in time until they reach a volume whose free energy is $10 \, k_{\rm B} T$ lower than the top of the barrier. We then proceed to pick $30$ points each from $10$ such reactive trajectories at random, yielding $300$ configurations on which the committor analysis is performed. Each step in the committor analysis of a given configuration consists of drawing random momenta corresponding to $296.4 \, {\rm K}$ and propagating the system in time until it reaches a boundary on either side of the barrier. For each configuration, we perform at least $10$ such steps until the error estimate $\sigma = \sqrt{p_{\rm B}(1 - p_{\rm B})/N} \leq 0.05$, where $N$ is the number of shots and $\sigma$ is the standard error in the committor assuming Gaussian statistics. 

\begin{figure}[h]
\centering
\includegraphics[width=0.35\textwidth]{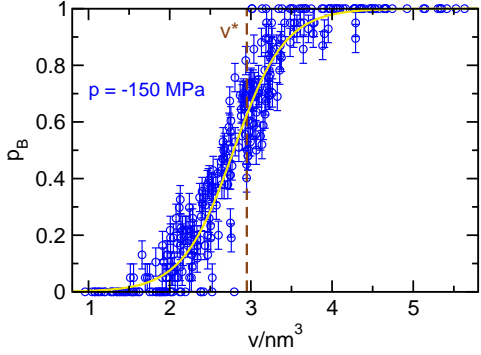}
\caption{
Committor $p_{\rm B}$ for configurations with a largest bubble of volume $v$. The configurations are randomly selected points along reactive trajectories at a pressure of $-150 \, {\rm MPa}$.
The yellow line is a fit using a hyperbolic tangent, error bars are $\pm \sigma$, and the dashed brown line indicates the location $v^\ast$ of the maximum in the free energy barrier shown in Fig.~1a. 
}\label{fig:committor}
\end{figure}

The result of this analysis, shown in Fig.~\ref{fig:committor}, reveals that the volume of the largest bubble in the system is a good reaction coordinate for cavitation in water. Higher values for the volume of the largest bubble correspond to higher committor probabilities and 
the spread of the data is moderate. 
As such, the volume of the largest bubble is suitable for the computation of rates via Eq.~(1) \cite{BerezhkovskiiSzaboJPhysChemB2013,LuVandenEijndenJCP2014}.
Further, the volume $v^\ast$ of the critical bubble obtained from free energy computations lies in the range of bubble volumes where transition states, i.e., configurations with $p_{\rm B} = 0.5$, are found \footnote{In general, the location of the maximum in $g(v)$ is not identical to $p_{\rm B} = 1/2$ even if $v$ parametrizes $p_{\rm B}$ perfectly, since $g(v)$ is not symmetric around $v^\ast$.}. 

In an effort to find correlations of $p_{\rm B}$ with other properties of the largest bubble, we investigated bubble asphericity, normalized surface to volume ratio, and the hydrogen bond structure at the interface, but none of these properties correlate with the committor in a statistically significant fashion.

\section{Surface free energy and curvature dependence of the surface tension}
In this section, we obtain a quantitative description of the cavitation free energy from CNT by examining the surface free energy, which allows to compare the free energetic cost of forming a liquid--vapor interface for different pressures. We then discuss the obtained curvature dependence of the surface tension that favors bubbles over droplets.

\begin{figure}[h]
\centering
 \includegraphics[width=0.35\textwidth]{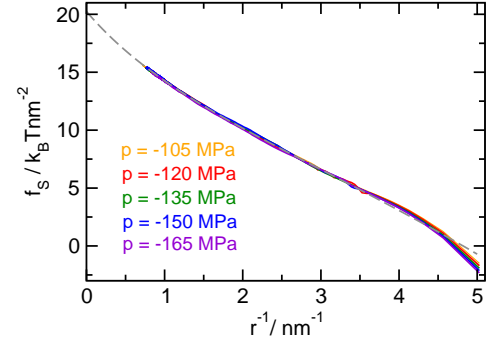}
 \caption{
Bubble surface free energy $f_{\rm s}$ as a function of inverse bubble radius $r^{-1} = (3 v / 4 \pi)^{-1/3}$.
As expected from theory, the bubble surface free energy does not depend on pressure (as will be discussed in the next section).
 The dashed grey line is a fit according to Eq.~(\ref{eq:fsfit}) for all pressures, where data in the range $0 < r^{-1} < 3.5\, {\rm nm^{-1}}$ were used for fitting.
 \label{fig:fs}}
\end{figure}

The surface free energy $f_{\rm s}$ is given by
\begin{equation}
 f_{\rm s} = a^{-1} (-k_{\rm B}T \ln [ v_0^2 \rho(v) ] - pv),
\end{equation}
where $a = (36 \pi v^2)^{1/3}$ is the surface area of a sphere with volume $v$, $v_0 = 1 \, {\rm nm^3}$ determines the unit of volume, and $\rho(v)$ is the equilibrium bubble density.
Here, $pv$ is the average mechanical work gained by expanding the system under tension when a bubble of volume $v$ is formed (see Methods);
by subtracting this contribution, we can compare the cost of forming a bubble in the metastable liquid at different pressures directly.
Furthermore, by fitting the surface free energy with a suitable functional form explained below, we elucidate the normalization constant $\rho_0$ relating the free energy $g(v)$ to the equilibrium bubble density $\rho(v)$ via $g(v) = -k_{\rm B} T \ln [ \rho(v)/\rho_0 ]$. 
Surface free energies for various tensions are shown as a function of inverse bubble radius $r^{-1} = (3 v/4 \pi)^{-1/3}$ in Fig.~\ref{fig:fs}.

Remarkably, the resulting surface free energy $f_{\rm s}$ is independent of pressure (a thermodynamic analysis of this behavior is provided in the subsequent section), except for very small bubbles. Consequently, we fit $f_{\rm s}$ for all pressures with the functional form 
\begin{equation}
\label{eq:fsfit}
f_{\rm s} = \gamma_0/( 1 + 2 \delta / r) + C/4 \pi r^2,
\end{equation}
which takes into account the curvature dependence of the surface tension via a Tolman-like correction. 
The fit yields $\gamma_0 = 20.24 \, k_{\rm B}T/{\rm nm^2}$, $\delta = 0.195 \, {\rm nm}$, and $C = -3.80 \, k_{\rm B}T$  (the result of the fit is indicated by the dashed grey line in Fig.~\ref{fig:fs}). Note that the constant $C$ is related to $\rho_0$ via $\rho_0 = \exp(\beta C)/v_0^2 = 0.022 \, {\rm nm^{-6}}$ and thus determines the normalization of the free energy $g(v)$ under the condition that the free energy of a bubble of vanishing size is zero, $\lim_{v \rightarrow 0} [g(v)] = 0$. Thus, we obtain all quantities needed to describe the free energy of cavitation, $g(v)$, in the framework of CNT using Eq.~(\ref{eq:tolman}).
 
The functional form of the free energy in Eq.~(\ref{eq:tolman}), whose parameters are obtained from the fit described above, is identical to the variant of CNT incorporating a curvature dependent surface tension proposed by Tolman \cite{Tolman} and as such it is tempting to identify the parameter $\delta$ with the Tolman length. 
However, a fundamental assumption required to obtain Eq.~(2) in the framework of the theory is that the radius $r$ of the bubble is large compared to the length $\delta$ \cite{RowlinsonWidom,TroesterBinderJCP2012} and thus the applicability of the Tolman formalism is questionable.
Yet, when studying cavitation in water at ambient temperature, this shortcoming is only relevant for the theoretical exercise of extracting the Tolman length, since Eq.~(2) describes the free energy of cavitation accurately over the range of volumes $v^\ast$ of critical bubbles at physically relevant conditions, i.e., conditions at which rates can be measured in experiment
(see Fig.~4).

The value of $\delta$ obtained from the fit shown in Fig.~1b is positive which indicates that the concave curvature of the interface decreases the surface tension $\gamma$.
In the literature, there are conflicting reports on the dependence of the surface tension on curvature in water:
Refs.~\cite{AzouziCaupinNaturePhys2013,JoswiakPetersJPCL2013,BruotCaupinPRL2016} find that the bubble is free energetically favored over the droplet, while Refs.~\cite{SedlmeierNetzJCP2012,VaikuntanathanGeisslerPRL2014,FactorovichScherlisJACS2014,WilhelmsenRegueraJCP2015,LauFordJCP2015} arrive at the opposite conclusion.
Notably, the value obtained from the fit for $\gamma_0$ is higher than the value $\gamma_0 = 17.09 \, k_{\rm B}T/{\rm nm^2}$ obtained by Vega and de Miguel \cite{VegaMiguelJCP2007} for the flat interface at ambient pressure.
This may be due to
a scenario similar to the behavior observed for very large spherical solutes in SPC/E water at ambient pressure. 
As shown in Ref.~\cite{SedlmeierNetzJCP2012}, the surface tension $\gamma(r)$ as a function of radius bends back to lower values, i.e., $\delta < 0$, for very large spherical cavities, which reconciles the estimate for $\gamma_0$ from the fit with the data for the flat interface at ambient pressure. In light of conflicting results on the sign of $\delta$ in water, further study is required to elucidate the influence of the chosen water model and biasing towards certain cavity shapes on the obtained value of $\delta$.

\section{Pressure dependence of the cavitation free energy}
The bubble surface free energy $f_{\rm s}(v)$, shown in Fig.~\ref{fig:fs}, is independent of pressure over the investigated pressure range. This results in bubble free energies $g(v)$ which only differ in the amount of mechanical work $pv$ gained by expanding the system under tension. In the following, we derive an analytical expression for the pressure dependence of $f_{\rm s}(v)$ and show that the change in free energy is negligible over a wide range of pressures.

For bubbles that do not occur spontaneously in the metastable liquid on the timescale of an unconstrained simulation, the bubble surface free energy
\begin{equation} \label{eq:fsurf2}
 f_{\rm s}(v) = \frac{1}{a} \left(-k_{\rm B}T \ln \left[ \frac{v_0^2 P(v)}{\langle V \rangle} \right] - pv \right) \, ,
\end{equation}
where $a = (36 \pi v^2 )^{1/3}$, $v_0$ is a constant that determines the unit of volume,
\begin{equation}
  P(v) = \frac{\int {\rm d}V \int \exp \left( - \beta \left[ H({\bf x}) + pV\right] \right) \delta \left[ v({\bf x}) - v\right] {\rm d}{\bf x}}{\int {\rm d}V \int \exp \left( - \beta \left[ H({\bf x}) + pV\right] \right) {\rm d}{\bf x}}
\end{equation}
is the equilibrium bubble probability density for the volume of the largest bubble,
and $\langle V \rangle$ is the average volume of the unconstrained metastable liquid. The bubble volume $v= \langle V \rangle_\xi - \langle V \rangle$ is the difference in system volume under a constraint $\xi$, i.e., a largest bubble of size $\xi$, and the metastable liquid, on average. In this work, the chosen constraint is the preliminary grid-based bubble volume estimate described in Ref.~\cite{ourfirstpaper}, but the following derivation is not limited to this specific bubble detection procedure.
The pressure derivative
\begin{equation}
 \frac{\partial f_{\rm s}(v)}{\partial p} = -a^{-1} 
  \frac{\partial }{\partial p} \left[ 
  k_{\rm B} T  
  \left( \ln \left[ v_0 P(v) \right] - \ln \left[ \frac{\langle V \rangle}{v_0} \right] \right) + pv
  \right] \label{eq:dfsdp} ,
\end{equation}
where we exploited the fact that $a^{-1} = (36 \pi v^2 )^{-1/3}$ is independent of pressure at the conditions studied here since $v$ is accurately reproduced by Eq.~(\ref{eq:fit2}) for all pressures (see Fig.~\ref{fig:Vvonv}). The pressure derivatives for the respective terms in the equation above are
\begin{alignat}{2}
-k_{\rm B} T \frac{\partial}{\partial p}  \ln \left[ v_0 P(v) \right] &= \langle V \rangle_\xi - \langle V \rangle &&= v \, ,\\
k_{\rm B} T \frac{\partial}{\partial p} \ln \left[ \frac{\langle V \rangle}{v_0} \right] &= \frac{k_{\rm B} T}{\langle V \rangle} \frac{\partial \langle V \rangle}{\partial p} &&= - k_{\rm B} T \kappa_{\rm T}
\, , \\
-\frac{\partial}{\partial p} pv &= -v - p \frac{\partial v}{\partial p} \, , \label{eq:dvdp}
\end{alignat}
where $\kappa_{\rm T}$ is the isothermal compressibility of the unconstrained metastable liquid. Over the investigated pressure range, the second term in Eq.(\ref{eq:dvdp}) vanishes since the change in $v(\xi) = \langle V \rangle_\xi - \langle V \rangle$ with pressure is negligible (see Fig.~\ref{fig:Vvonv}).
The resulting expression for the change in surface free energy with pressure is
\begin{equation}
 \frac{\partial f_{\rm s}(v)}{\partial p} = -a^{-1} k_{\rm B} T \kappa_{\rm T} \, . \label{eq:dfsdpfinal}
\end{equation}
Under the conditions studied here, i.e., when $\partial v / \partial p$ vanishes, the only $v$-dependent contribution remaining from Eq.(\ref{eq:dfsdp}) is $a^{-1}$. Consequently, the change in free energy with pressure is limited to a contribution to the normalization constant $\rho_0$ that relates the cavitation free energy $g(v)$ to the equilibrium bubble density $\rho(v)$ via $g(v) = -k_{\rm B} T \ln [ \rho(v)/\rho_0 ]$.

In practice, the change in surface free energy given by Eq.~(\ref{eq:dfsdpfinal}) is small due to the low compressibility of water. 
We compute the difference $\Delta f_{\rm s}(v) = f_{\rm s}(v,-165 \, {\rm MPa}) - f_{\rm s}(v,-105 \, {\rm MPa})$ in surface free energy between the highest $p = -165 \, {\rm MPa}$ and lowest $p = -105 \, {\rm MPa}$ tension investigated via
\begin{equation}
\Delta f_{\rm s}(v) = a^{-1} k_{\rm B} T \int_{-105\, {\rm MPa}}^{-165\, {\rm MPa}} \frac{1}{\langle V(p) \rangle} \frac{\partial \langle V(p) \rangle}{\partial p} {\rm d}p \, .
\end{equation}
\begin{figure}[h]
\centering
 \includegraphics[width=0.35\textwidth]{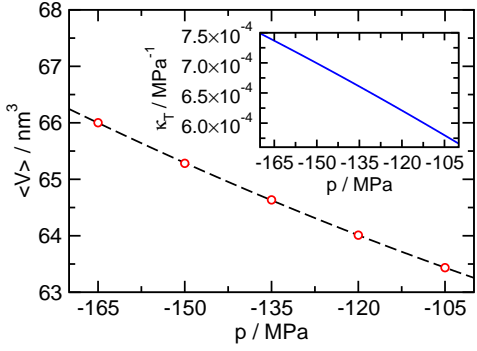}
 \caption{
 Average volume $\langle V \rangle$ of the unconstrained metastable liquid as a function of pressure. The dashed black line indicates a polynomial fit to second order. {\it Inset:} Isothermal compressibility $\kappa_{\rm T} = -\langle V \rangle^{-1} \partial \langle V \rangle/ \partial p$, where $\partial \langle V \rangle/ \partial p$ was computed by taking the pressure derivative of the fit in the main plot.
 \label{fig:compressibility}}
\end{figure}
Here, we compute the derivative $\partial \langle V(p) \rangle / \partial p$ by fitting a second order polynomial to the average volume $\langle V \rangle$ of the metastable liquid for different tensions and taking its derivative (see Fig.~\ref{fig:compressibility}).
The resulting estimate for $\Delta f_{\rm s}(v) \approx a^{-1} 0.04 \, k_{\rm B} T$ is smaller than the statistical uncertainty in Fig.~\ref{fig:fvonv}b.

\section{Curvature-corrected bubble dynamics}

Following the same procedure as in the main text, we obtain the cavitation rate estimate from CNT including a curvature-dependent surface tension $\gamma (r) = \gamma_0 / ( 1 + 2 \delta / r)$. 
First, we show that the functional form of the diffusivity $D(v)$ determined from the Rayleigh--Plesset (RP) equation does not change when the influence of curvature is taken into account. We then obtain the diffusivity and rate expression using the curvature-corrected estimates for the volume $v^\ast$ of the critical bubble and the curvature $\omega$ of the free energy barrier. 

Since the surface tension $\gamma(r)$ depends on the radius $r$ explicitly, we cast the RP equation in terms of the bubble radius for simplicity.
The RP equation with a curvature dependent surface tension $\gamma (r)$ reads
\begin{equation}
 m r \ddot{r} + \frac{3 m \dot{r}^2}{2} = p_{\rm b} - p - \frac{2 \gamma (r)}{r} - \gamma '(r) - \frac{4 \eta \dot{r}}{r} \, .
\end{equation}
Note that the equation above is equivalent to Eq.~(\ref{eq:rp}), describing the time evolution of the bubble radius $r$ instead of its volume $v$, with an additional term $\gamma ' (r) = {\rm d} \gamma / {\rm d} r$ that accounts for the change of the surface tension $\gamma(r)$ with bubble radius $r$.
Neglecting the inertial terms on the left hand side and the pressure $p_{\rm b}$ inside the bubble leads to
\begin{equation}
 \dot{r} = - \frac{r}{4 \eta} \left[ p + \frac{2 \gamma}{r} + \gamma '(r) \right] = - \frac{1}{\Gamma(r)} \frac{{\rm d}g(r)}{{\rm d} r} \, ,
\end{equation}
where the effective force
\begin{align}
- \frac{{\rm d}g(r)}{{\rm d} r} &= - \frac{{\rm d}}{{\rm d} r} \left[ 4 \pi r^2 \gamma(r) + \frac{4 \pi r^3}{3} p \right] \nonumber \\
&= -4 \pi r^2 \left[ \frac{2 \gamma(r)}{r} + \gamma ' (r) + p \right] .
\end{align}
The computed friction $\Gamma (r) = 16 \pi \eta r$ has the same form as the Stokes friction of a sphere dragged through a viscous liquid, but differs from it by a numerical factor.
Analogous to the derivation for non-corrected CNT we include thermal noise $F(t) = \sqrt{2 k_{\rm B} T / \Gamma(r)} \xi(t)$ in the RP equation and 
obtain the diffusivity $D(r) = k_{\rm B} T / 16 \pi \eta r$ via the Einstein relation. 
By casting the resulting Langevin equation in terms of the bubble volume $v$, we compute the diffusivity $D(v) = 3 k_{\rm B} T v / 4 \eta$ that has the same functional form as in the case of a constant surface tension. Consequently, the diffusivity at the top of the barrier $D(v^\ast$) differs from Eq.~(\ref{eq:DRP}) only in the estimate for $v^\ast$.

The volume of the critical bubble $v^\ast$ in curvature-corrected CNT is given by
\begin{equation}
  v^\ast = \frac{4 \pi}{3} \left( \frac{\gamma_0}{\lvert p \rvert} \right)^3 \left( 1 - \frac{4 \delta}{r_0^\ast} + \sqrt{1 + \frac{4 \delta}{r_0^\ast}} \right)^3 , 
 \label{eq:suppvstar}
\end{equation}
where $r_0^\ast = 2 \gamma_0/\lvert p \rvert$ is the estimate for the radius of the critical bubble from uncorrected CNT.
For the highest and lowest tension studied here, Eq.~(\ref{eq:suppvstar}) predicts critical bubble volumes $v^\ast$ which are reduced by a factor of $0.47$ and $0.65$ from the uncorrected CNT estimate $v_0^\ast = 32 \pi \gamma_0^3 / 3 \lvert p \rvert^3$, respectively.
Provided that $\delta$ is small compared to the radius $r_0^\ast$ of the critical bubble in uncorrected CNT,
i.e., $\sqrt{1 + 4 \delta/r_0^\ast} \approx 1 + 2 \delta/r_0^\ast$, the above equation can be rewritten as
\begin{align}
v^\ast &\approx v_0^\ast - 4 \pi \delta r_0^{\ast 2} \nonumber
\end{align}
when quadratic and higher order terms of $\delta p/\gamma_0$ are neglected.
Inserting Eq.~(\ref{eq:suppvstar}) yields the estimate for the diffusivity $D(v^\ast)$ on top of the barrier
\begin{align}
 D(v^\ast) &= \frac{3 k_{\rm B} T v^\ast}{4 \eta} \label{eq:suppD} \\
&\approx \frac{3 k_{\rm B} T v_0^\ast}{4 \eta} - 3 \pi k_{\rm B} T \delta r_0^{\ast 2} \, .
\end{align}

In order to obtain an estimate for the cavitation rate $J$ one requires the curvature $\omega = \sqrt{\vert {\rm d^2} g(v^\ast) / {\rm d} v^2 \vert }$ at the top $g(v^\ast)$ of the free energy barrier:
\begin{equation}
  \omega = \sqrt{\frac{\gamma_0}{2 \pi}} \frac{1}{r^{\ast 2}} \sqrt{\frac{1 + \frac{4 \delta}{r^\ast}}{(1 + \frac{2 \delta}{r^\ast} )^3}} \, .
\end{equation}
Inserting $r^\ast = (3 v^\ast /(4 \pi))^{1/3}$ yields an estimate for $\omega(v^\ast)$ which is similar to that obtained using a constant surface tension, $\omega_0(v_0^\ast) = p^2/\sqrt{32 \pi \gamma_0^3}$, differing by a factor of $1.27$ and $1.15$ for the highest and lowest tension investigated, respectively.

Inserting the quantities computed above into Eq.~(\ref{eq:rates2}) gives the rate estimate when the curvature dependence of the surface tension is taken into account and the correct value of $\rho_0$ is used. Note that in the evaluation of the data presented in the main text, the exact expressions in Equs.~(\ref{eq:suppvstar}) and (\ref{eq:suppD}) were used.

\section{Viscosity of water under tension}
To estimate the diffusion coefficient $D$ of a bubble in the framework of the RP equation
\begin{equation}\label{AppEq:D}
 D(v) = \frac{3 k_{\rm B} T v}{4 \eta} ,
\end{equation}
the viscosity of water under tension is required. 
The shear viscosity $\eta$ of a fluid can be computed by use of the Green--Kubo relation 
\cite{Tuckerman}
\begin{equation}
 \eta = \frac{V}{k_{\rm B} T} \int_0^\infty \langle P_{\alpha \beta}(0) P_{\alpha \beta}(t) \rangle \, {\rm d}t \, .
\end{equation}
Here, $\langle P_{\alpha \beta}(0) P_{\alpha \beta}(t) \rangle$ is the equilibrium autocorrelation function of the 
five independent components $P_{\alpha \beta}$ of the pressure tensor, namely $P_{\rm xy}$, $P_{\rm xz}$, $P_{\rm yz}$, $(P_{\rm xx} - P_{\rm yy})/2$, and $(P_{\rm yy} - P_{\rm zz})/2$. 
 By averaging over the autocorrelation functions of these five independent components we maximise the data harvested from each trajectory \cite{NevinsSperaMolSim2007}. 
 
We compute the pressure tensor $P_{\alpha \beta}$ as a function of pressure at various fixed volumes corresponding to average pressures in the range of interest and at a temperature $T = 296.4 \, {\rm K}$ in molecular dynamics simulations over a time of $6 \, {\rm ns}$.
The autocorrelation functions are evaluated from the power spectrum with fast Fourier transforms according to the Wiener--Khintchine theorem,
up to a time of $20 \, {\rm ps}$.
After averaging the autocorrelation functions over all off-diagonal pressure tensor components $P_{\alpha \beta}$, the integration is carried out numerically and the estimate for $\eta$ is obtained by fitting the emerging plateau for long times with a constant.
\begin{figure}[h]
\centering
 \includegraphics[width=0.35\textwidth]{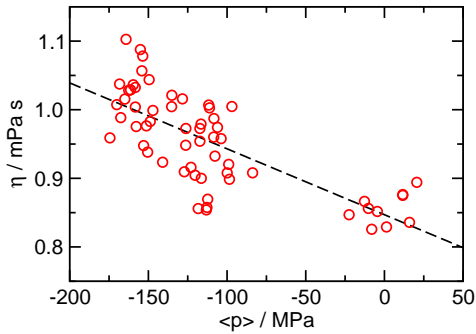}
 \caption{
 Viscosity $\eta$ as a function of pressure $p$. Computations were performed at constant volume, the pressure $\langle p \rangle$ is the canonical average for each data point.
 The dashed black line is a linear fit to the data.
 \label{fig:eta}}
\end{figure}
The resulting estimates for $\eta$ are shown in Fig.~\ref{fig:eta}. The shear viscosity increases with tension, consistent with the behavior at positive pressures reported in Ref.~\cite{GonzalezAbascalJCP2010}.
Due to the large scatter in the data and in absence of prior knowledge about the functional form of $\eta(p)$, we use a linear fit on the data. Doing so results in good agreement with the literature value $0.855 \, {\rm mPa \, s}$ for TIP4P/2005 water from Ref.~\cite{GonzalezAbascalJCP2010} at ambient pressure. 
While the statistical error in the viscosity $\eta$ is relatively large, and hence the functional dependence of $p$ cannot be reliably extracted from the data, we stress that the viscosity is not the only pressure dependent factor entering Eq.~(\ref{AppEq:D}). In particular, when using Eq.~(\ref{AppEq:D}) with the Kramers equation (see Eq.~(\ref{eq:rates2})), 
the change in $\eta$ with pressure is very small compared to the change in $v^\ast$. Thus, the exact scaling behavior of $\eta$ will not significantly influence the estimates for the diffusion constant obtained from the RP equation in conjunction with CNT (let alone the estimate for the rates which are dominated by the change in free energy with pressure).

\section{Comparison of the obtained rates to experimental data}
To put the cavitation pressure presented in Fig.~\ref{fig:rates} into context, we discuss its relation to experimental results obtained from different setups.
Our estimate for the cavitation pressure, obtained from the cavitation rates calculated for typical experimental conditions, $p_{\rm cav} \approx -126 \, {\rm MPa}$, can help disentangle the experimental situation. As discussed before \cite{CaupinHerbertCRPhys2006,CaupinJNonCrystSol2015}, experiments fall in two major groups. On the one hand, a set of very different techniques reach similar $p_{\rm cav}$ around $-30 \, {\rm MPa}$. On the other hand, only one technique (water inclusions in quartz) seems to reach beyond $-100 \, {\rm MPa}$ \cite{GreenAngellScience1990,ZhengAngellScience1991,AlvarengaBodnarJCP1993,AzouziCaupinNaturePhys2013,MercuryShmulovichNATO2014,PallaresCaupinPNAS2014}. A first possible explanation for this discrepancy in measured cavitation pressures is that the pressure reported for the inclusion experiments is not correct, because an extrapolated equation of state is used to infer $p_{\rm cav}$ from the fluid density in the inclusion and the temperature $T_{\rm cav}$ at which cavitation occurs. This explanation was excluded based on direct measurement of the cavitation density \cite{DavittCaupinEPL2010} and, more recently, by a direct experimental determination of the pressure reached in inclusions \cite{PallaresCaupinPCCP2016}. The latter work provides an equation of state down to $-95 \, {\rm MPa}$ at around $325 \, {\rm K}$; a short extrapolation then confirms that pressures close to $-120 \, {\rm MPa}$ have been reached the experiments discussed in Ref.~\cite{AzouziCaupinNaturePhys2013}. 

Two different scenarios can explain the discrepancy between experiments \cite{DavittCaupinEPL2010}: (i) either homogeneous cavitation occurs in water close to $-30 \, {\rm MPa}$, and water in inclusions is stabilized by some unknown mechanism, or (ii) homogeneous cavitation occurs close to $-120 \, {\rm MPa}$, and, apart from the inclusion work, nucleation occurs heterogeneously in experiments at lower tensions because of ubiquitous impurities. 
The cavitation rates obtained in the present work based on molecular simulation of pristine water, which result in $p_{\rm cav} \approx -126 \, {\rm MPa}$,
support the second scenario.
This result is in good agreement with density functional theory calculations \cite{CaupinPRE2005}, while CNT with the prefactor employed in the rate prediction shown in Fig.~\ref{fig:rates} yields a stronger tension of $p_{\rm cav} \approx -176 \, {\rm MPa}$.
Based on theoretical predictions, the second scenario is therefore more likely, although the nature of the impurities inducing cavitation at $-30 \, {\rm MPa}$ is still unclear. 
A recent shock pulse study \cite{StanDeckerJPhysChemLett2016} proposes that, for extremely fast pressure ramps, homogeneous cavitation beyond $-100 \, {\rm MPa}$ could occur concurrently with heterogeneous cavitation because the bubbles from heterogeneous nucleation forming at around $-30 \, {\rm MPa}$ do not have enough time to grow sufficiently to release the tension in the system. 

Finally, we note that heterogeneous nucleation also occurs in some inclusions. Indeed, in a given quartz sample containing many inclusions with similar liquid density, a wide range of $p_{\rm cav}$ has been observed \cite{ZhengAngellScience1991,MercuryShmulovichNATO2014}. Analyzing the details of the nucleation statistics in a given inclusion \cite{AzouziCaupinNaturePhys2013} clearly shows that the scatter of nucleation temperatures for a given inclusion is fully consistent with nucleation theory: cavitation is a stochastic event depending on the thermodynamic conditions, the sample volume, and its cooling rate. However, the distribution of nucleation temperatures in a given inclusion is quite narrow, around $5 \, {\rm K}$, one order of magnitude less than the broad range observed between different inclusions with the same density. It must be concluded that heterogeneous nucleation (on dissolved impurities or surface defects) is responsible for the scatter of $p_{\rm cav}$ in inclusions. However, in the inclusions with the largest $p_{\rm cav}$, it is assumed that nucleation occurs homogeneously. It is of course possible that further experiments would find inclusions exhibiting an even more negative $p_{\rm cav}$. However the numerous experiments already performed and the consistent trend observed for the most negative $p_{\rm cav}$ vs. density suggests that the homogeneous nucleation limit has been reached. The value obtained for $p_{\rm cav}$ in the present work supports that this is indeed the case.

\end{document}